# Depth resolution optimized sputter depth profiling of a polycrystalline Al layer

Uwe Scheithauer, ptB, 82008 Unterhaching, Germany
E-Mail: scht.ptb@t-online.de; scht.uhg@mail.de
Internet: orcid.org/0000-0002-4776-0678; www.researchgate.net/profile/Uwe_Scheithauer

**Abstract:**

Depth profiling of thin films by inert gas sputtering is one of the most important applications of Auger electron spectroscopy and X-ray photoelectron spectroscopy. Such an analysis monitors the elemental in-depth composition of the thin film system and controls the quality of the manufacturing process. A challenging task of a sputter depth profile measurement is the determination of the interface contaminations. Interface contaminations are only detectable if the depth profile data were measured with a sufficient depth resolution. Interface contaminations are crucial, because they strongly influence the mechanical and electrical properties of the whole thin film system.

The depth resolution of unidirectional sputter depth profiling for polycrystalline samples is limited by the sputter yield differences attributed to grains having different crystalline orientations relative to the incoming ion beam. Therefore depth resolution optimized sputter depth profiling of polycrystalline thin films requires dedicated experimental approaches. If the sample is rotated during ion sputtering the depth profile is recorded with a better depth resolution because the ion impact direction relative to the grains orientation varies. If depth profiling can be performed on a single grain only, the depth resolution is improved, too. These approaches are applied to the depth profiling and interface contamination analysis of a 5000 nm thick polycrystalline Al layer. An O contamination at the interface of the Al to the Si substrate was detected utilizing these high depth resolution depth profiling methods.

## 1. Introduction

In-depth profiling of thin film systems is one of the most important applications of Auger electron spectroscopy (AES) and X-ray photoelectron Spectroscopy (XPS). Applying these techniques the sample surface is eroded by ion bombardment ("sputtering") usually using inert gas ions. The residual surface is analyzed after each sputter step. As a function of sputter time the depth distributions of the elements are recorded [1, 2].

The samples crystallinity has a great influence when polycrystalline thin films are analyzed. For some grains the impinging ions move parallel to open crystal axes. Due to this channeling the energy and momentum of the sputter ions is transferred to the sample in greater sample depth. For these grains the sputtering is less effective compared to those where the sputter cascade occurs directly underneath the sample surface. The sputter yield differences result in a poor depth resolution of the sputter depth profile if the measured signal integrates laterally over several grains of different orientation [3] and the ion impact angle relative to the sample remains fixed. This unidirectional depth profiling represents the conventional sputter depth profiling procedure.

In this paper a 5000 nm thick polycrystalline Al layer, which was deposited on a Si wafer, was analyzed applying unidirectional depth profiling, sample rotation depth profiling [4, 5] and single grain depth profiling [6]. It is demonstrated that the depth resolution at the interface between Al and Si substrate improves significantly by sample rotation depth profiling and especially by single grain depth profiling. Thanks to this better depth resolution an interface contamination becomes detectable.

## 2. Instrumentation

An XPS microprobe and an AES microprobe were used for the sputter depth profile measurements. The AES microprobe has a much better lateral resolution, because an electron beam can easily be focused by electrical and magnetic fields.

The XPS microprobe is a Physical Electronics XPS Quantum 2000. The Quantum 2000 achieves its spatial resolution by the combination of a fine-focused electron beam generating the X-rays on a water cooled Al anode and an elliptical mirror quartz monochromator, which monochromatizes and refocuses the X-rays to the sample surface. Nominal beam diameters between 10 µm and 200 µm are adjustable. Details of the instruments design and performance are discussed elsewhere [7-14]. For sputter depth profiling the instrument is equipped with a differentially pumped $Ar^+$ ion gun. $Ar^+$ ion energies between 0 and 5 keV are selectable. Ion energies between 250 eV and 5 keV are used for sputtering. For sample surface charge compensation low energy $Ar^+$ ions are utilized. The incoming X-rays are parallel to the surface normal if the samples are flat mounted on the standard sample holder. In this geometrical situation, the mean geometrical energy analyzer take off axis and the differentially pumped $Ar^+$ ion gun are oriented ~ 45° relative to the sample surface normal. With a low voltage electron flood gun sample surface charging can be compensated, too. The samples are mounted on a 75 x 75 $mm^2$ sample holder. It is transferred into the main analysis chamber via a separate turbo pumped vacuum chamber.

A Physical Electronics PHI 680 Auger microprobe was used for the high lateral resolution measurements [15]. The PHI 680 microprobe, an instrument with a hot field electron emitter [16], has a lateral resolution of ~ 30 nm at optimum. A lateral resolution of ~ 50 ... 70 nm is achievable under analytical working conditions operating the instrument with a higher electron current. Electron energies between 1 and 25 keV are selectable. The electron beam is guided coaxially to the central axis of the cylindrical mirror analyzer. The Auger microprobe is equipped with a differentially pumped $Ar^+$ ion sputter gun. Sputter ion energies between 0 and 5 keV are selectable. Ion energies between 250 eV and 5 keV are used for sputtering. For sample surface charge compensation low energy $Ar^+$ ions are utilized. If the electron beam hits the sample under an angle of 30° relative to the sample surface normal, which is the standard working condition, the incoming $Ar^+$ ions impinge onto the surface under an angle of 55° relative to the surface normal. The instrument is equipped with a sample transfer system. The samples are mounted on sample holders, which are transferred into the main analysis chamber via a separate turbo pumped vacuum chamber.

All data evaluation was done by an improved version of the PHI software MultiPak 6.1 [17]. It has been improved by a module, which enables an evaluation of measured depth profile data utilizing non-linear least square (= nlls) fitting by internal reference spectra. During the fit procedure each spectrum in each sputter depth is fitted by a linear combination of these reference spectra. Small peak energy shifts of the reference spectra compensate slight in-depth energy variations due to sample charging. The first advantage of nlls is that this fitting procedure improves the signal-to-noise ratio, because the nlls fitting uses all recorded data. Second, peak overlays are deconvoluted this way. Especially, chemical depth distributions of a single element are evaluated. Differently shaped spectra, which represent different chemical bondings of a single element, are measured during the sputter depth profiling.

## 3. Sputtering of Polycrystalline Samples

$Ar^+$ ions with energy of 2 keV are used for the sputter depth profile measurements presented here. A sputter ion impact angle of ~ 45° relative to the surface normal is used in the XPS Microprobe and an angle of ~ 55° is used in the AES microprobe. Since there is only a low probability at these impact angles that a sample atom will be sputtered away by a single collision with an incoming ion, the momentum necessary to eject a surface atom is transferred to the atom in almost all cases via a sputter cascade.

Due to this, beside other parameters, the sputter efficiency depends on the sample depth where the energy and momentum is transferred to the sample. At this point the sample crystallinity has a great influence. When sputtering polycrystalline samples, for some grains the incoming ions move parallel to open crystalline axes, channeling occurs [18, 19]. Hence the energy and momentum of an incoming ion is transferred to the sample in greater sample depth. For these grains the sputtering is less effective as for those where the sputter cascade occurs directly beneath the sample surface. This effect results in a poor depth resolution of the sputter depth profile if the measured signal integrates over several grains of different crystalline orientation.

An interface contamination is smeared out proportional to the depth resolution of a sputter depth profile measurement. One monolayer on top of a sample will contribute ~ 10 to 30 at% to the signal of an Auger or XPS measurement, dependant on the contaminations and the substrates elemental composition. At optimum the elemental detection limits of an electron spectroscopy measurement are in the range between 0.1 to 1 at% dependent on the elements sensitivity factor. Using these hand waving arguments: A one monolayer interface contamination is no longer detectable if it is smeared out over hundreds of monolayers. Therefore a good depth resolution of a sputter depth profile measurement is an indispensable prerequisite for the detection of interface contaminations.

Here an Al thin film layer sample is analyzed by unidirectional, rotational and single grain depth profiling to demonstrate the advantages of the advanced approaches. In case of unidirectional depth profiling the impact angle of the sputter ions remains constant. During rotational depth profiling the sample rotates around its surface normal during sputtering to avoid permanent ion channeling on single grains since the ion impact angle relative to the grains crystal axes varies. During a single grain measurement the analysis area is restricted to one grain only. Single grain depth profiling is a promising approach if the grain sizes of the analyzed layer are larger than the film thickness [6].

## 4. Experimental Details

Depth profiles of a 5000 nm thick Al layer were measured applying different sputter depth profile approaches. The Al layer was deposited on a Si wafer. Secondary electron (SE) images were taken after the depth profile measurements. The SE images show that the Al layer is polycrystalline (fig. 2, 4 & 6). Many of the crystals have lateral dimensions which are larger than the layer thickness.

A unidirectional depth profile and a sample rotation depth profile were recorded using the XPS microprobe. A 200 µm X-ray beam was used. The samples were sputtered by 2 keV $Ar^+$ ions. The $Ar^+$ beam was rastered by electrostatic deflection to get homogenous sample erosion. The AES microprobe was utilized for depth profile measurements on single grains of the Al layer, because it has a much better lateral resolution than the XPS microprobe. The AES microprobe was operated with a 10 kV / 20 nA electron beam. The sample was sputtered by

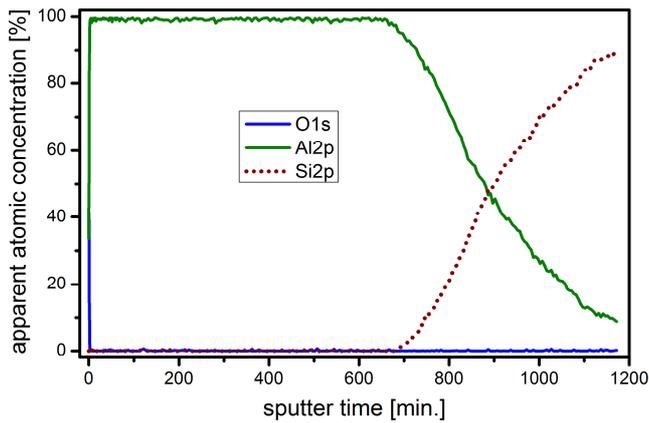

Fig. 1: unidirectional XPS depth profile
interface depth resolution:
Δt/t (Al 75% ->25%) = 25.4 [%]
2 keV Ar$^+$, ion impact angle: ~ 45° to surface normal

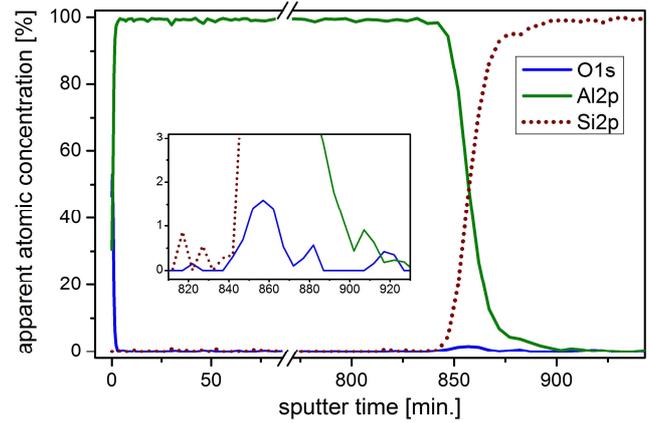

Fig. 3: XPS depth profile measured with sample rotation during sputtering
interface depth resolution:
Δt/t (Al 75% ->25%) = 1.14 [%]
2 keV Ar$^+$, ion impact angle: ~ 45° to surface normal
The insert shows the signals at the interface.

2 keV Ar$^+$ ions. For homogenous sample erosion the Ar$^+$ beam was rastered electrostatically.

## 5. Experimental Results

Figure 1 shows the result of a unidirectional XPS depth profile measurement. Since the X-ray beam diameter is larger than the grains dimension of the polycrystalline Al layer this depth profile measurement integrates over many grains. The depth resolution of the interface is rather poor. This is due to the different crystalline orientation of the single grains relative to the Ar$^+$ ion incidence direction. Therefore the sputter yields of the grains differ. The depth resolution is 24.5 % if the Al signal decrease from 75% to 25% is used for estimation. A sputter depth resolution of this magnitude for unidirectional depth profiling is typical for an Al metallization used in microelectronic devices [3]. At the end of the measurement, after a sputter time of 1170 min., ~ 9.5 at% of Al are detected. SE images were recorded after sputtering (see fig. 2). These images depict that the remaining Al signal is due to Al grains, which had not been completely sputtered.

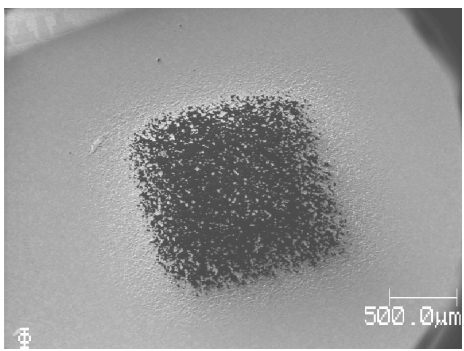
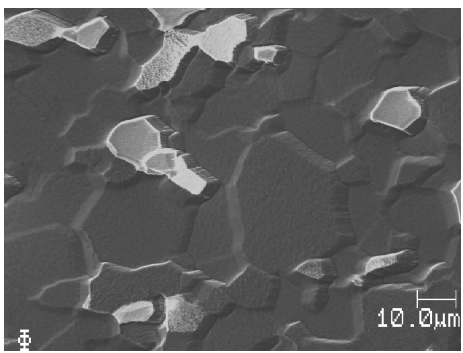

Fig. 2: SE images of the sample after a unidirectional XPS depth profile measurement

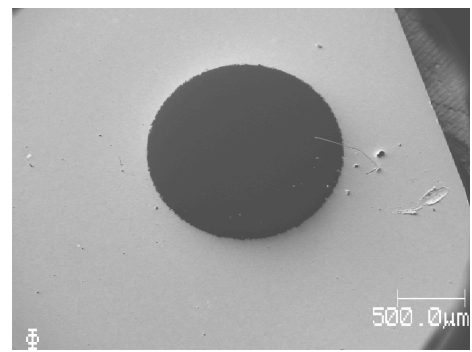
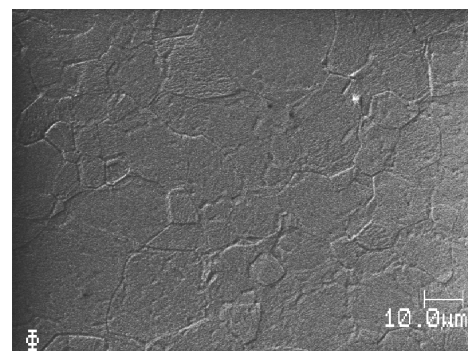

Fig. 4: SE images of the sample after rotational XPS depth profile measurement

The rotation of the sample around its surface normal during sample sputtering results in a much better depth resolution of the measurement (fig.3). At the end of the measurement no Al signal is detected. The interface depth resolution Δt/t is 1.14%. The insert of figure 3 indicates an O contamination at the interface. The SE images (fig. 4), which were recorded after sputtering, show a homogenous surface without any remaining Al.

Figure 5 reports the results of single grain depth profiles recorded on 4 different grains using the AES microprobe. Only the Al and Si data at the interfaces are plotted. Depended on the different grain orientation relative to the impinging $Ar^+$ ions and the corresponding different sputter yields the interfaces are reached after different sputter time. The sputter time is normalized by the fastest sputtered grain. Sputter time differences and sputter yield differences, respectively, up to a factor of ~ 2 are estimated for the 4 different oriented grains. The SE image (fig. 6), which was recorded

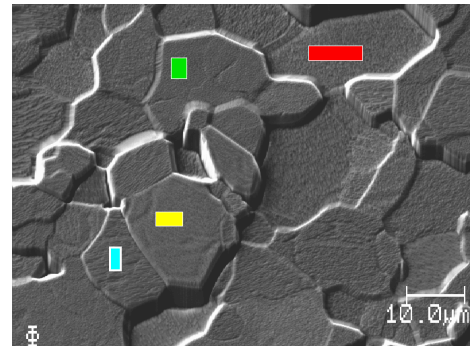

Fig. 6:  SE image of the sample after AES depth profile measurement
4 different measurement areas are marked after depth profiling, depicts the analysis areas. The height steps between the different grains visualize the sputter yield differences. In the lower part of figure 5 the estimated depth resolutions are plotted. For the 4 grains the interface depth resolution Δt/t varies between 0.67% and 0.82%. The depth resolution is slightly better then the depth resolution of the sample rotation depth profiling (fig. 3). Figure 7 shows another AES single grain depth profile measurement. For this measurement an ultimate depth resolution of 0.5% is estimated. The $O_{KLL}$ concentration was elaborated by nlls fitting using the $O_{KLL}$ spectrum of the surface as internal reference spectrum. The insert of figure 7 indicates an O contamination at the interface.

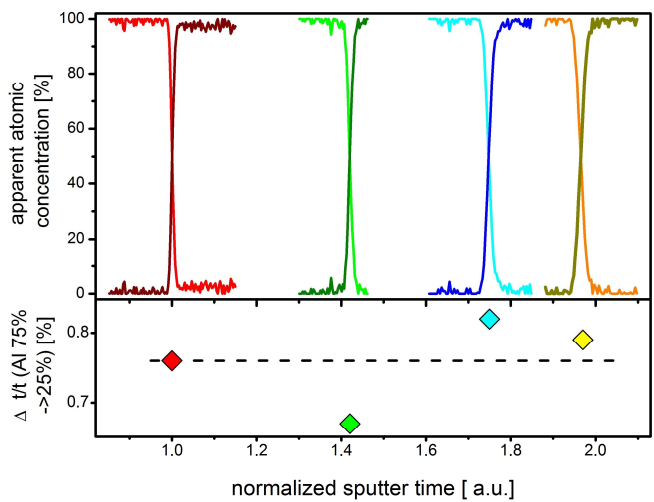

Fig. 5:  AES depth profiles measured at 4 different single grains
2 keV $Ar^+$, ion impact angle: ~ 55° to surface normal

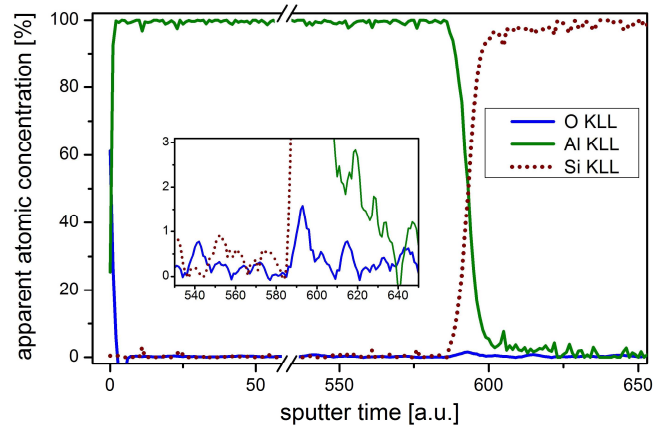

Fig. 7:  AES single grain depth profile measurement
interface depth resolution:
Δt/t (Al 75% ->25%) = 0.5 [%]
2 keV $Ar^+$, ion impact angle: ~ 55° to surface normal
The insert shows the signals at the interface.

## 6. Conclusion

A 5000 nm polycrystalline Al layer and its interface to the Si substrate were analyzed by XPS and AES sputter depth profile measurements. The advanced approaches using sample rotation or single grain depth profiling, respectively, measure the elemental depth distributions with a significantly higher depth resolution than the unidirectional depth profiling. The unidirectional depth profiling results in a depth resolution of Δt/t = 24.5 %, because it integrates laterally over many grains with different orientation relative to the impinging sputter ions. Sample rotation during sputtering improves the depth resolution to Δt/t = 1.14 %. In this measurement the signals integrates over many grains, too. But due to the sample rotation the $Ar^+$ ion impact angle relative to the grains crystal orientation changes continuously. During the rotation the sputter yield of every single grain varies and the sputter yields of the different grains equalize. For the five single grain depth profile measurements an ultimate depth resolution Δt/t between 0.5 % and 0.82 % was estimated. Since the analysis area is restricted to a single grain, the different sputter yields of grains with different orientation relative to the impinging $Ar^+$ ions have no significant impact to the depth resolution. Only the sputter time, which is needed to reach the interface, varies.

The enhanced depth resolution of the rotational and single grain depth profile measurements enables the detection of an O contamination at the interface between the 5000 nm Al layer and the Si substrate. These measurements demonstrate exemplarily the benefit of advance sputter depth profiling techniques. Only this way interface contaminations become detectable. Interface contaminations have a great impact to the mechanical and electrical properties of a thin film system. Therefore the detection of interface contaminations is essential for judging on the quality of polycrystalline multilayer thin films as used for microelectronic purposes, for instance.

### Acknowledgement

The measurements were done utilizing an XPS microprobe Quantum 2000 and an AES microprobe PHI 680, respectively, installed at Siemens AG, Munich, Germany. I acknowledge the permission of the Siemens AG to use the measurement results here. For fruitful discussions and suggestions I would like to express my thanks to my colleagues. Special thanks also to Gabi for text editing.

.